\documentclass[12pt,preprint]{aastex}  

\slugcomment{Version of \today}
\shortauthors{Smith \& Balona}
\shorttitle{A Study of HD\,110432} 
\begin{document} 

\newcommand{\iue}{{\it IUE}}
\newcommand{\hd}{HD\,110432}
\newcommand{\gam}{$\gamma$\,Cas}

\title{The Remarkable Be Star HD110432 (BZ Cru)}

\author{Myron A. Smith} 
\affil{Catholic University of America,\\
        3700 San Martin Dr.,
        Baltimore, MD 21218;           \\
        msmith@stsci.edu }

\and
\author{Luis Balona\altaffilmark{1}}
\affil{South African Astronomical Observatory,\\
        P. O. Box 9, 
        Observatory 7935, South Africa;           \\
        lab@saao.ac.za }

\begin{abstract}

\hd~  (B1e) has gained considerable recent attention because it is
a hard, variable X-ray source with local absorption and also because 
its optical spectrum is affected by an extensive Be disk. 
From time-serial echelle data obtained over two weeks during 2005 January 
and February, we have discovered several remarkable characteristics in the
star's optical spectrum. The line profiles show rapid variations on some 
nights which can most likely be attributed to irregularly occurring and
short-lived {\it migrating subfeatures.} Such features have been found 
in spectra of \gam~ and AB\,Dor, two stars for which it is believed 
magnetic fields force circumstellar clouds to corotate over the star's 
surface. The star's optical spectrum also exhibits a number of mainly 
Fe\,II and He\,I emission features with double-lobed profiles typical of 
an optically thin circumstellar disk viewed nearly edge-on.  Using spectral 
synthesis techniques for the January data, we find that its temperature 
and column density are close to 9,800\,K and roughly 
3$\times$10$^{22}$ cm$^{-2}$. 
Its projected disk size covers a remarkably large 100 stellar areas, and
the emitting volume resides at a surprisingly large distance of 1\,A.U. 
from the star. Surprisingly,  we also find that the absorption wings of 
the strongest optical and UV lines in the spectrum extend to at least ${\pm 
1000}$ km\,s$^{-1}$, even though the rotational velocity is 300--400 
km\,s$^{-1}$.  We are unable to find a satisfactory explanation for these 
extreme line broadenings. Otherwise, \hd~ and \gam~ share similarly peculiar 
X-ray and optical characteristics. These include as high X-ray temperature, 
erratic X-ray variability on timescales of a few hours, optical metallic 
emission lines, and submigrating features in optical line profiles. Because 
of these similarities, we suggest that \hd~ is a member of a select new 
class of ``\gam~ analogs." 

\end{abstract}
 

\clearpage

\section{Introduction}
\label{int}

  HD\,110432 (BZ\,Cru = HR\,4830, m$_v$ = 5.3) is a bright variable 
late-Oe to Be-type star (Houk \& Cowley 1975) located in the open cluster 
NGC\,4609 (Feinstein \& Marraco 1979) and beyond the southern Coalsack. 
In recent years this star has attracted considerable attention, resulting 
in several investigations in the X-ray, ultraviolet, and optical wavelength
regimes. In each regime, the star has been found to be unusual. 
The optical spectrum is distinguished by Balmer and Fe\,II 
lines having double-lobed emission profiles (Slettebak 1982), which is
an unusual attribute for classical Be stars. Because the optical
spectrum does not show forbidden Fe lines, nebular lines, or
features due to a cool companion, it fails the definition of a B[e] star.  
The star's substantially reddened colors and linear polarization have 
made it an important candidate for ultraviolet and optical for studies 
of interstellar extinction. In this respect the star's reddening 
is unusual and suggests that its observed flux distribution has been 
altered by circumstellar matter (Meyer \& Savage 1981). 

 In their classic X-ray survey of potential Be X-ray binary systems, 
Torrej\'on \& Orr (2001, hereafter TO01) found that \hd~ is a hot thermal 
X-ray source. Its light curve, extracted from BeppoSax spectra, 
does not show rapid pulses ($\le$10$^3$ sec) characteristic of a Be X-ray 
pulsar system, transients, or alternating extended high and low states 
that are typical of most Galactic X-ray Be stars (and CV systems).
These authors found that the spectrum can be fit 
with a MEKAL model with a temperature of $kT$ = 10.55\,keV or with
a power law fit with a photon index $\alpha$ = 1.35 and
cut-off energy E$_c$  = 16.69 keV. The power-law fit may be improved
by the addition of Fe 6.7 and 6.9 keV emission lines.  However, since
the presence of these lines is consistent with the slope of the X-ray
continuum, a thermal interpretation is to be preferred. The absorption 
column density from the attenuation of soft X-ray continuum flux in this
model is about 1.08$\times10^{22}$ cm$^{-2}$, or fully a factor of ten
higher than the ISM column densities inferred from ultraviolet atomic 
and molecular hydrogen lines (Rachford et al. 2001). 

 TO01 extracted a 5-hour light curve from their BeppoSax observations 
and discovered that it was characterized by a 4-hour modulation.  TO01 
supposed that the variation in this light curve is periodic and ascribed 
it to a 4-hour rotation of a hypothetical white dwarf with a spotted 
surface in a binary system. Because modulations on this timescale are common
in the light curve of \gam~ (Robinson, Smith, \& Henry 2002, hereafter
RSH02) and since an X-ray temperature of  $\approx$10$^{8}$K (10.55\,keV)
is also characteristic, of \gam~ (e.g., Owens et al. 1999, Smith et al. 
2004), the TO01 study suggested that follow-up observations on \hd~ 
would be productive to establish whether it showed the optical light 
curve and spectroscopic attributes of \gam~ as well.

 Our discussion in this paper proceeds by first 
describing the observational and modeling tools ($\S$\ref{tls})
used in the analysis and reviewing the known fundamental parameters 
of the \hd~ ($\S$\ref{fund}). 
We then explore the star's known X-ray and optical light curves ($\S$4)
and the properties of archival UV resonance line ($\S$5) and newly obtained 
optical spectra ($\S$6).  In $\S$\ref{concl} we outline the likely properties 
of members of a ``\gam" class of stars.

\section{Observations and Modeling Tools}
\label{tls}

\subsection{Optical data} 
\label{obsn}

   In this study we obtain photometric and spectroscopic observations
of \hd~ to investigate whether its optical properties
might be related to the X-ray variability. The photometric observations
were carried out by the observing staff of the 0.5-m
telescope at African Astronomical Observatory's Sutherland station
during 2002 February 6--May 7 in the Cousins $V$ and $B$ filters.
The comparison stars used for
these observations were two late-type B stars, HR 4736 and HR 4944.
All three stars were observed through the same neutral density filter
and standardized with repect to ``E-region" UBVRI photometric standards.

  To investigate the properties of the disk and the star's optical
variability over short time scales, we undertook a high resolution
spectroscopic campaign during February  2005. Using the fiber-fed
{\it Giraffe} spectrograph attached to the 1.9-m telescope of the South
African Astronomical Observatory (SAAO). This echelle spectrograph is a copy 
of the MUSICOS spectrograph of the Telescope Bernard Lyot at Pic du Midi
Observatory (Baudrand \& Bohm 1992). The configuration places 50 spectral
orders on a  1024 $\times$ 1024 TEK CCD detector covering the range 
$\lambda\lambda$4270--6800. The spectral resolution was 32,000, and 
the dispersion ranged from 0.06--0.09\,\AA\,pixel$^{-1}$. 
A Th--Ar arc lamp was used to provide a wavelength calibration, with arc
spectra being taken at regular intervals to minimize instrumental drift.
Flatfielding was accomplished by illuminating the camera with uniform
light provided by a tungsten filament lamp and diffusing screen. 
Echelle blaze functions were determined by measuring the response across each
order when the fiber was illuminated by the tungsten lamp. This procedure
was largely (but not completely) effective in rectifying the continuum.
The data reduction was carried out using the ``SPEC2" spectroscopic data 
reduction package, as used by most other observers of this instrument.

 We obtained 142 spectra during on the nights of 2005 January 26 --February 1.
Dr. David Laney kindly obtained 61 additional spectra for us on 
February 22--March 1, 2005. Typically we obtained a signal-to-noise 
ratio of 50 per pixel in an exposure time of 10--20 minutes. 
Nine of the fifteen nights were suitable for a search for
{\it migrating subfeatures,} which are discussed below, according 
to our (somewhat arbitrary) criterion of at least 10 observations 
distributed over at least three hours on a given night.

In order to place the continuum in an objective a manner, we first 
constructed a synthetic spectrum with similar effective temperature and
gravity.  The lines were then broadened for the appropriate rotational
velocity.  We then found the ratio between the observed and synthetic
spectrum and selected those points which are on the continuum, according 
to the synthetic spectrum.  This step involves filtering and smoothing.
Finally, a low-order polynomial (usually of degree 3) was fitted to the
observed points on the spectrum deemed to be on the continuum.  This
procedure is completely automatic and has been used in all observations of
early-type stars made with {\it Giraffe}.  Comparison between the observed
spectrum and the synthetic spectrum is always very good except for the
region around the Balmer lines.

\subsection{Spectral synthesis modeling }
\label{syncir}

  In $\S$\ref{emissp} we will discuss results of a number of line 
syntheses we performed using the {\it SYNSPEC}
code, version 48, written by Hubeny, Lanz, and Jeffery (1994). 
As input for these solutions, we used the atomic data from the 
line library (Kurucz 1990) and a standard model atmosphere (Kurucz
1990) having parameters $T_{\rm eff}$ = 25,000\,K, log\,g = 4, and
$\xi$ = 2 km\,s$^{-1}$. Our synthesized spectra were computed in 
steps of 0.01 \AA.~ We assumed standard abundances for these
computations and convolved the result with an Unsold 
``rotation function" using computed continuum intensities determined for
a grid of 20 $\mu$ angles from a vertical direction in the atmosphere.
To compute the effect of a circumstellar ``cloud" or disk on the 
spectrum we used the output from our {\it SYNSPEC} calculation as input
for a computation of cloud absorption or emission by the {\it CIRCUS} 
program (Hubeny \& Heap 1996).  

In both these programs we used a working value of
V$_{rot}$sin\,i = 300 km\,s$^{-1}$ as determined in $\S$\ref{fund}. 
However, this value need not be known
precisely because the narrower emission line contributions can be easily
separated from the underlying broad photospheric line absorptions.
We ran {\it CIRCUS} in its LTE 
reemission mode, that is, by evaluating the radiative transfer relations 
assuming a locally defined excitation and ionization temperature. 
Other important disk physical parameters used to compute the equivalent
widths of lines formed in the disk are its column density, projected 
disk area, microturbulence, and chemical composition (assumed herein to be
solar). A volumetric particle density of 1$\times$10$^{12}$ cm$^{-3}$ was 
chosen from the disk model for \gam~ (Millar \& Marlborough 1998; their 
Table\,1). This density corresponds to a point one scale height out of 
the disk plane and about 1R$_*$ from the star's surface.  
{\it CIRCUS} handles as many as three clouds with independent velocity 
and thermodynamic parameters.  Except in one case, we imposed the 
condition that our model disks were homogeneous.  
For the exception as noted below, we utilized this capability and 
constructed a two-temperature disk model.

  In many such applications, one has to make educated guesses to set the 
most important parameter, the disk temperature.
In this paper we will be able to provide an accurate
determination of the temperature by the simultaneous presence of iron and
helium line strengths in the disk spectrum. The temperature guess drives 
our values of other parameters. Another important determination is the 
regime of optical depth (thick or thin?) through which the lines are 
observed. The equivalent width ratios of iron lines having different 
atomic $gf$ values enable one to estimate the optical depths and the hence 
column densities of the disk region. However, we should emphasize in
advance that while these properties define the total volume of the emitting 
gas, they do not define its distance from the exciting star or the internal 
density distribution. These conditions are better defined by separate 
kinematic arguments.

\section{Fundamental physical properties of HD\,110432} 
\label{fund}

  Observations of \hd's optical and UV spectrum suggest a spectral type in
the range of B0.5 to B2  (Hiltner et al. 1969, Codina et al. 1984, Ballereau,
Chauville, \& Zorec 1995). Although previous distance estimates have run 
as high as 430${\pm 60}$ pc (Codina et al. 1984), {\it Hipparcos} 
parallaxes have led to a more modest value of 300 ${\pm 50}$ pc (Perryman 
1997). 

  Surveys of OB stars have determined that the UV reddening curve of 
\hd~ is peculiar (e.g., Meyer \& Savage 1981). In their summary of
E(B-V) reddening results for this star in the literature,
Rachford et al. (2001) concluded that a ``low" value of  E(B\,-\,V) 
= 0.40 magnitudes results if the star's parameters lies within the 
spectral type and class ranges of B0--B2 and V-III, respectively, 
if the {\it Hipparcos} parallax is correct within its errors,
and if the reddening toward the star is representative of the average 
Galactic relation. One can reconcile this low 
E(B\,-\,V) with the value of 0.51 found by Meyer \& Savage if one
attributes a reddening difference $\Delta$E(B\,-\,V) $\approx$0.11
to a circumstellar nebula or disk near the star.

  Zorec, Fr\'emat, \& Cidale (2005) have recently calibrated the Balmer 
jump criteria ($D,$ $\lambda_1$) they measure in 97 OB stars to 
current-generation model atmospheres. This work allowed them to assign
values of $T_{\rm eff}$ = 22,510\,K and log\,g = 3.9 to \hd. However,
we have found that the strengths of certain UV temperature-sensitive
lines indicate a slightly higher temperature.  For example, the He\,II 
$\lambda$1640 line is almost as strong (90--95\%) as in the \gam~ 
spectrum. In addition, Codina et al (see their Fig.\,1) determined
parameters T$_{\rm eff}$ = 25,000\,K and log\,g = 3.5 from a 
UV-to-near-IR fitting of the star's spectral energy distribution with
models computed from model atmospheres. 
We have adopted the Codina et al. value for our line syntheses. This
is slightly cooler than most estimates of the $T_{\rm eff}$ for \gam, 
and so we adopt a spectral type of B1\,IV.

  The radial velocity of HD110432 is uncertain because its few strong 
absorption lines are very broad.  An examination of the literature shows 
determinations ranging from +9 to +35 km\,s$^{-1}$. Because of this
scatter, the star is generally noted as being variable in velocity, and
this is probably due to emission of its blue lines for at least
some decades. For
example, we note the comment by Thackeray, Tritton \& Walker (1973; TTW73): 
``The H emission lines are very strong, being visible at least as far as 
H9. The velocities are mainly based on Balmer emission. He absorption is 
faintly visible (~B8\,V) and Fe\,II emission suspected." We have adopted 
for this study the most recently determined value of +9 km\,s$^{-1}$ by 
TTW73.

  According to the literature, the rotational velocity of \hd~ has 
been obtained from either the $\lambda\lambda$4471-4481 optical region or
the UV resonance lines. In the first case, various optical studies have 
attempted to measure the $V_{rot}sin\,i$ from the He\,I 4471\AA\ line.
(The often used nearby Mg\,II 4481\AA\ line is hopelessly blended in
this star's spectrum.) These studies have given results in the range of 
300--380 km\,s$^{-1}$ (Slettebak 1982, Ballereau, Chauville, \& Zorec  
1995). However, the latter authors believed that this line was probably 
mutilated at the time of their observation (Zorec 2005). 
In light of our new data (see
$\S$\ref{brabs}), we believe that any velocities derived from fitting this
this line are open to question. Codina et al. (1984) were able to fit 
the C\,IV and Si\,IV resonance lines with a broadening parameter of
of 360 km\,s$^{-1}$ that they attributed to rotation. However, as we will
develop below, there appears to be an issue as to whether the components 
these authors identified are photospheric. In view of these uncertainties, 
we display in Figure\,\ref{fit11} the $\lambda\lambda$1115--40 region of 
the FUSE spectrum, as obtained from the MAST archives. In this figure
we have fit the observed spectrum with a synthetic spectrum computed with 
{\it SYNSPEC} using a rotational parameter $V_{rot}sin\,i$ = 
300 km\,s$^{-1}$ with an 5\% emission component from a warm (20,000\,K) 
circumstellar cloud.  The {\it FUSE} spectra of \hd~ and \gam~ in this
spectral region exhibit
essentially the same broadening. We have also performed a similar 
exercise for the fitting of the He\,II 1640 \AA\ line, as shown 
in Figure\,\ref{he2fit}, and for the Fe\,III line-rich region of  
$\lambda\lambda$1910-1930 (not shown). We found a good fit to the
line profiles with  $V_{rot}sin\,i$ = 300 km\,s$^{-1}$ in both cases.
While these rotational velocity determinations are internally consistent, 
they are at the low end of the 300--380 km\,s$^{-1}$ range quoted above 
from fitting of optical lines. The disparity with our result is an example 
of a well-known tendency of UV line analyses to result in lower rotational 
velocities than analyses of optical lines. We believe that optical 
determinations should be preferred over our own because the nearby continuum
can be placed with more confidence. 
For example, we find that a rise in
the fluxes in the $\lambda\lambda$1637--1638 region, probably a true
continuum reference window, is not easily matched in our syntheses of
spectra of several other early-type Be stars.
In addition, it is likely that gravity
darkening in these very rapidly rotating stars deemphasizes the contribution
of lines formed near the equator. Such lines would be formed at higher latitude
regions where the total range of local rotational velocities is smaller.
A convincing demonstration of this principle was shown in the case of the
rapdily rotating star $\zeta$\,Oph by Reid et al. (1993). These authors noted
the absence of line profile variations due to equatorially confined 
nonradial pulsations in high-excitation lines of He\,II and N\,III, while
lines of less excited ions showed variations from the equatorial mode.
The most important result to take away from this work is that the
velocity is probably {\it at least} 300 km\,s$^{-1}$. Whatever the precise
value is, Fig.\,\ref{fit11} shows that it is likely nearly the same
as the velocity of \gam.

\section{Light Curve Modulations }
\label{xr}

\subsection{The X-ray light curve of Torrej\'on and Orr }

 By kind permission of J. Torrej\'on, we have reproduced the 
BeppoSax light curve published by TO01 as Figure\,\ref{xrhd}. Their
light curve shows a dip in X-ray flux followed about two hours later
by a rise to the initial flux.  Although TO01 suggested that 
the variation in this light curve is periodic, a single cycle leaves
open the possibility that its variations can be characterized in 
other ways. To show how another interpretation is possible, we depict 
in Figure\,\ref{xrgam} an {\it RXTE} light curve of \gam~ over a
comparable stretch of time. This particular time sequence was taken from
that shown by Figure\,2 of Robinson \& Smith (2000), and it is centered at
about 10 UT on 1998 November 24. Superimposed on this figure is the
sine curve fit obtained by TO01 for the \hd~ data, adjusted
for count rate. Clearly, the TO01 sine wave for \hd~ is a good fit
to the \gam~ variations. Yet, a glance at the full data train in the
Robinson \& Smith figure makes it obvious that ``sinusoidal
variation" in the \gam~ curve is actually part of a series of a 
seemingly random meandering pattern that suggest that the appearance 
of a periodicity is illusory. From this comparison the temporal
characteristics of the light curves of \hd~ and \gam~ are essentially
the same.

\subsection{Optical cycles? }

  Because both the X-ray temperature and variability properties of
\hd~ are reminiscent of \gam, we decided to look for other possible 
similarities in terms of published X-ray/optical correlations. 
  
  We began our optical study of HD\,110432 by a photometric monitoring
campaign in 2002. This followed the report by  RSH02,
since validated by Smith, Henry, \& Vishniac (2006, hereafter SHV06) 
of nonrepeating cycles in the light curve of \gam.~ These
variations have a characteristic length of 60--90 days and have a
full amplitude of $\simeq$3\%. This amplitude increases slightly from
the Johnson $B$ to $V$ filter bandpass.  Interestingly, RSH02 also 
found that the X-ray flux of \gam~ varies by a factor of three 
over timescales of months. Moreover, these authors found that the
X-ray and optical variations correlate very well with one another,
a fact substantiated by more recent optical and X-ray observations
reported by SHV06.
Since they are both cyclical, they do not correlate with the star's
binary orbital phase.\footnote{\gam~ is a nearly circular binary with a 
period of 204 days, according to Harmanec et al. (2000) and Miroshnichenko, 
Bjorkman, \& Krugov (2002).  the mass and evolutionary status of the 
secondary of \gam~ are unknown.  The binary status of \hd~ is currently 
unknown.} The cyclical and ``reddish" nature of the optical variations
led RSH02 to propose that the X-ray variations in \gam~ are generated
by a magnetic dynamo excited in the Be star's (decretion) disk.

 To see whether \hd~ exhibits optical characteristics similar to the optical
modulations that in \gam~ also correlate with X-ray variations, we 
undertook the photometric monitoring campaign described in $\S$\ref{obsn}.
The results of our 2002 campaign in Sutherland showed an identical
variability pattern in both $V$ and $B$ filter data. We exhibit our
results for the $V$ band in Figure\,\ref{ltcol}a.
This figure shows that after the start of the monitoring period 
the optical magnitude of \hd~ faded rapidly by 3--3${\frac 12}$\%. 
After a gap in monitoring of nearly two months, the star once again regained 
its initial brightness. At the end of the observing season, the star's
faintness faded to its mean level. This variability is typical of a cyclical 
behavior in Be stars. If this particular cycle is representative of the 
variability of this star, then the cycle amplitude, waveform, and length 
are all similar to the optical cycles in \gam. In addition, according to the
preponderance of scatter in the upper left of the constant-color relation
of Fig.\,\ref{ltcol}b, the star's brightest changes were greater during 
these variations in the $V$ band than the $B$ band. These optical properties 
are similar to those RSH02 found for \gam.

\section{Variability of the strong ultraviolet resonance lines }

Codina et al. (1984) have discussed the strong wind absorptions in the 
blue wings of the resonance lines of C\,IV, N\,V, and Si\,IV at one 
epoch in 1981. This absorption is accentuated by the presence of DACs 
(Discrete Absorption Features) in the C\,IV and Si\,IV doublets. 
Figure\,\ref{civ} displays a sample of three {\it IUE} ({\it Short 
Wavelength Prime}) spectra in the neighborhood of the C\,IV doublet 
and Si\,IV line. These examples include the 1981 SWP spectrum 
discussed by Codina et al. (shown as the dotted line) as well as a 
mean of two spectra from 1986 and one spectrum from 1991. The 
1986 and 1991 spectra indicate that the wind absorption in the C\,IV
and Si\,IV (though not shown, also N\,V ) doublets had been comparatively 
weak in 1981. In all epochs a rich array of DACs were present in the
C\,IV lines, and indeed their detailed structure can be quite different 
The wind absorption can be discerned in the C\,IV $\lambda$1548 line 
out to -1800 km\,s$^{-1}$. Coincidentally or not, this is also the 
absorption edge limit found in resonance line spectra of $\gamma$\,Cas 
(e.g., Cranmer, Smith, \& Robinson 2000, hereafter 
CSR00). Another similarity with the $\gamma$\,Cas spectrum is that 
the Si\,IV DAC typically has a single
component located at about -1100${\pm 200}$ km\,s$^{-1}$ (cf.
Doazan et al. 1987, Smith, Robinson, \& Corbet 1998).

  In contrast to the very stable He\,II $\lambda$1640 discussed above,
the resonance lines of Si\,IV, C\,IV, and even N\,V exhibit extensive
variations even at low (positive and negative) velocities. The
position line minimum can vary noticeably on a timescale of 2--3 days, 
suggesting that much of the central core is formed in the wind at some
times. Moreover, as indicated by the dotted and dashed arrows in 
Fig.\,\ref{civ} the minima occurs at different velocities in the various 
ions at any given time. This leads us to wonder whether the feature,
blueshifted to -100 km\,s$^{-1}$, and utilized by Codina et al. (1984) 
in the 1981 spectrum, gives a fair measure of the photospheric velocity
field. A detailed check of the Si\,IV profiles during
an intensive monitoring 2-day campaign in 1986 suggests that the central
core and low velocities, out to -300 km\,s$^{-1}$, develop and stagnate
at irregular intervals of several hours.  For example, the absorptions 
in this region remain static for the spectra given by SWP\,27967, 27970, 
and 27972, strengthen for SWP\,27980, 27991, and 27997, and remain 
static again by the time of the SWP\,28004 exposure. Similar patterns
may be present, but less clearly in the blue wing of the leading member
of the C\,IV and N\,V doublet. This type of activity is reminiscent of
the recurrence of wind features on the timescale of many Be stars'
rotational periods. In the particular case of $\gamma$\,Cas these patterns
can be associated with particular regions close to the Be star's surface
which in turn emit high X-ray fluxes and absorb UV continuum flux,
respectively (Smith \&
Robinson, hereafter SR99; Cranmer, Smith, \& Robinson 2000).

\section{Optical high-resolution spectroscopy }
\label{opt}

\subsection{Migrating subfeatures }

\subsubsection{The migrating subfeatures in line profiles of \gam }

  In 1988 Yang, Ninkov, \& Walker discovered the existence of a
peculiar new class of traveling bumps in the line profiles of \gam. 
These ``migrating subfeatures" ({\it msf}) form in the blue wing 
and move at a rate exceeding the expected star's surface rotation rate. 
They generally form in loose groups, and although they are usually 
present they can be occasionally absent on any given night. Often an
{\it msf} can form or disappear (or both) during the time it moves 
across the line profile, and typically their lifetimes are 3 hours 
or less (Smith 1995).  Although Yang, Nikov, \& Walker observed {\it msf} 
in the optical line profiles of $\gamma$\,Cas, these
features have since been observed in high signal-to-noise time-serial 
ultraviolet spectra  obtained by the Goddard High Resolution
Spectrograph, formerly attached to the {\it Hubble Space Telescope} (SR99). 
Spectral line syntheses of the ultraviolet 
data suggests that the {\it msf} are likely the absorptions caused
by intervening clouds anchored to the surface by putative magnetic fields 
(Smith, Robinson, \& Hatzes 1998). 
In these higher quality data, it becomes apparent that faint {\it msf} are 
numerous and can be identified with lifetimes even as short as about 1 
hour (SR99).  Until this study, 
{\it msf} have been reliably reported in the line profiles of \gam~ 
and only one other star, AB\,Dor (Cameron Collier \& Robinson 1989). 
The latter is an active, magnetic pre-main sequence K dwarf (e.g., Hussain
et al. 2000). An additional discovery of {\it msf} was announced for the 
B5 star HR\,1011 (Smith 1996), but adequate follow-up observations 
not yet been made. Additional
authors have pointed to the existence of probably short-lived 
clumps of circumstellar gas close to Be stars (Smith \& Polidan 1993, Smith 
1995, Peters 1998, Zorec, Fr\'emat, \& Hubert 2001), which may be related 
to but do not technically qualify as {\it msf.}

\subsubsection{Discovery of migrating subfeatures in line profiles of \hd }

To investigate whether the line profiles of \hd~ exhibit any type of rapid
variability, we undertook the spectroscopic campaigns in 2005 January 
and February described above. From the resulting time sequences we computed 
grayscales from the difference spectra of each nightly mean spectrum. 
In these computations we first removed cosmic ray events and isolated 
bright pixels by correcting them to the mean flux of their neighbors. 
We also removed minor artificial continuum undulations from the initially 
computed difference spectra by removing undulations found in fifth 
degree polynomial fits. 

In our grayscale results for six nights (the fifth of our seven nights in 
January was largely clouded out), we found clear {\it msf} patterns for two 
nights. These results are presented in Figure\,\ref{hdmsf}. This patterns 
were slightly weaker on the third night and weaker still on the fourth and 
sixth night. By the seventh night no {\it msf} pattern could be detected. 
Similarly, of three nights in February having 10 or more observations, 
{\it msf} were present on two of them and not on a third.
  Note on January 27th (see Fig.\,\ref{hdmsf}) the clear three narrow dark
features running diagonally toward the upper right part of the grayscale. 
On January 25/26 one can see that the end of a first, and the beginning 
of a fourth, {\it msf}-like feature are present. In addition, we estimate
an acceleration rate of +100${\pm 10}$
km\,s$^{-1}$hr$^{-1}$ in the {\it msf} features in Fig.\,\ref{hdmsf}. 
Within the errors this value is the same as the rate +92\,--\,95 
km\,s$^{-1}$hr$^{-1}$ for the {\it msf} found in line profiles of \gam~ 
(e.g., Smith 1995, SR99). 

  The presence of small co-rotating clouds anchored to a rapidly 
rotating star seems to be a necessary condition for inferring the presence 
of magnetic fields in rapidly rotating stars. Clearly, this inference
cannot be readily made from zeeman observations.
For this reason any claim of detection of {\it msf} calls for a discussion
of mechanisms that might confuse this interpretation. A reminiscent
type of spectral pattern is the collection of traveling ``bumps" 
produced on line profiles by high-degree nonradial pulsations (NRP). 
Although NRP are generally found in cooler spectral types than B1, 
they have been observed in late-O main sequence stars $\zeta$\,Oph (e.g., 
Reid et al. 1993, Kambe 1997, Balona \& Kambe, 1999, Walker et al. 2005) 
and HD\,93521 (Howarth et al. 1998), and for the B0.5\,IV-III star 
$\epsilon$\,Per (Gies et al. 1999). 
The acceleration values of the migrating NRP bumps from NRPs in the
line profiles of these massive stars cases tend to be clumped either
at high values, e.g. near 150 km\,s$^{-1}$hr$^{-1}$ for $\zeta$\,Oph, 
and HD\,93521, or at values much lower than the corotation rate, at 
$\approx$20 km\,s$^{-1}$hr$^{-1}$ for $\epsilon$\,Per.

   Superficially, the line profile striations in these stars seem 
similar to the {\it msf} in line profiles of \gam. However, in general 
the dark striations caused by NRP and by intervening corotating clouds 
differ in several respects:

\noindent {\it i)} the spacings of {\it msf} are irregular, such that 
the time intervals between successive transits can change, while for 
monoperiodic NRP the bumps are evenly spaced.

\noindent {\it ii)} the absorptions constituting the {\it msf} are confined
to small wavelength intervals compared to the spaces between them. 
In contrast, the waveforms of monoperiodic NRP are nearly sinusoidal. The
dark and light bands in the grayscale depiction have nearly the same width.

\noindent {\it iii)} {\it msf} can be present on any one night and 
disappear or appear only sparsely on a second night. The available data 
suggest that both they and their associated X-ray activity centers have 
limited lifetimes of only one to a few days (Robinson \& Smith 2000,
RSH02). In contrast, the NRP bumps are present every night, and their 
visibility varies over short timescales only because of the interference 
of bump systems arising from different modes of comparable strength.

\noindent {\it iv)}  the {\it msf}~ in \hd~ and \gam~ exhibit accelerations 
of 92--100 km\,s$^{-1}$hr$^{-1}$. This is close to the value for 
rotational migration of a spot on the star's surface, and yet also consistent
with migrations from corotating clouds near the surface (e.g., Smith 
1995, SR99). In the NRP paradigm an acceleration rate must be consistent
with the intrinsic periods suggested by NRP theory. Such periods are
thought to be of the order of the timescale in the metal bump zones in
which they are driven (e.g., Pamyatynkh 1998), which are several hours
for an early B-type main sequence star. In contrast, an acceleration
rate consistent with surface corotation is by definition infinitely long 
in the corotating frame and therefore disallowed by NRP theory. 
From this argument, the observed acceleration rate in \hd~ is consistent 
only with the corotation scenario.
    
\noindent {\it v)} {\it msf} seldom last during a full transit from 
the blue to the red wing whereas NRP bumps due so as a rule. 
High signal-to-noise observations of the UV lines of \gam~ 
(Smith, Robinson, \& Corbet 1998, SR99) indicate that weak {\it msf} are
ubiquitous in spectral regions with a dense packing of lines. 
While current optical spectra have permitted the identification of the 
stronger ones that last as long as 3 hours, higher quality {\it HST/GHRS}
data shows that the more numerous faint ones exit with lifetimes even 
as short as about 1 hour (SR99).

\noindent {\it vi)} the acceleration of bumps produced by NRP in a
line profile increases toward the edges of the line profile due to the
foreshortening of the distance traversed by the waves toward the 
stellar line. A nice illustration of this effect is provided by
Peters \& Gies (2005) in their grayscale depicting NRP in $\pi$\,Aqr.
In contrast, the {\it msf} enter/exit the profile edges with an
acceleration closer to the value at the center of the profile. This
is because the velocity vector of an elevated cloud has not completely 
rotated from a traverse to a radial orientation by the time the cloud 
has moved off the projected limb.

   All of these {\it msf} attributes found for \gam~ are met in our 
observations of \hd. However, the first two points listed above are 
risky discriminators between NRP and co-rotating clouds in themselves. 
For NRP stars with multiple high-degree modes, like $\zeta$\,Oph, the 
bands {\it can} sometimes appear at irregular intervals and appear as
nonsinusoidal waveforms.
 Point {\it iii,} the occasional absence of features, is more difficult 
for NRP to mimic, although it could be reproduced if one sampled the 
line profiles at unlucky times. Even so, it is difficult for us to 
conceive that this coming and going of features could occur in the NRP 
interpretation both in our January and February monitorings.

  Points {\it iv} and {\it v} constitute somewhat stronger arguments that
these features are {\it msf.}  According to point {\it iv,} the
observed acceleration rate in \hd~ is consistent only with a corotation
rate, presumably magnetically constrained, of a structure located near 
the star's surface. In addition, the discovery of co-rotating clouds 
{\it close} to a star's surface is preferred because of the rapid fall 
off (1/$r^ {3}$) in magnetic field strength from the surface.  
The discovery of a star with an even larger magnetic field strength 
that could confine plasma at higher elevations would be correspondingly 
less probable.

  Point {\it v} is likewise persuasive.
   From a perusal the data of Reid et al. (1993), the convenient 
observing timespan needed to differentiate between NRPs and {\it msf} 
on any particular night seems to be of the order of 7--8 hours. In this 
timescale, one can begin to decide whether the features are ``permanent," 
and that they do not disappear either because the intrinsic disturbance 
dissipates 
or because an elevated cloud responsible for it moves off the background
stellar disk.  Our observing durations are too short to provide this 
type of conclusive test. However, some of our features do disappear or 
appear suddenly near the middle of the profile, and this suggests short 
lifetimes. 

  Collier \& Robinson (1989) have pointed out that the rotation of the
velocity of a corotating cloud can be utilized to determine the cloud's
height around the star (point {\it vi}). In our data the limited 
signal-to-noise ratio and short lifetimes of the profile features limit 
our ability to make a good estimate. However, one can say that there is 
little evidence for a dramatic acceleration of the features near the 
${\pm v_r}$ edges in our grayscales. This fact precludes the formation 
of these features on or within about 0.3R$_{*}$ of the surface.

   We conclude this discussion by stating that it is probable but not proven 
that the time-dependent striations we observe are induced by magnetically
confined clouds. Clearly we cannot make firm conclusions in this regard 
based only on a few nights of data. The identification of {\it msf} 
properties in \gam~ itself was the result of several investigations of 
many UV and optical lines.

\subsection{ The optical emission spectrum } 
\label{echvr}

  Because the instrumental configurations were static during our January
and February observing runs, we could form average spectra from our 
two datasets. The coadded spectra for both epochs exhibited the same 
peculiarly shaped emission profiles noted by Slettebak (1982). 
The only prominent absorption lines in our spectra are the first
three Balmer lines and the helium He\,I 4471 \AA,~ 5876 \AA,~ and 6678
\AA\ lines.  Of these features, only $\lambda$4471 is present entirely
in absorption. Other light-element lines have absorption wings and 
the same double-lobed emission cores as the metallic lines. We
will discuss the absorption features further in $\S$\ref{brabs}.

\subsubsection{Identification and profiles of emission features}

  \hd~ has an optical spectrum contains numerous permitted hydrogen,
helium, and metallic lines in 
emission. For this reason already it belongs to a subset of ``classical 
Be stars"\footnote{The spectra of classical Be stars have Balmer emission 
lines produced by decretion disks. These disks are formed neither by
protostellar collapse or binary mass transfer.} with this characteristic. 
Also, the profiles are marked by a flat central 
emission plateau flanked by stronger $V$ and $R$ lobes. This type of
double-peaked emission profile is common in a small subset of classical
Be stars and has been modeled by Hanuschik (1988) and Hanuschik et al. 
(1996).  The double-lobed profile is a signature of an optically thin 
Keplerian disk with little or no radial expansion or contraction. 
The star-disk system is seen nearly edge-on (precisely edge-on 
systems show a narrow central ``shell" component not present in our 
spectra of HD\,110432). Moreover, the two peaks are generally
slightly unequal. Over time they track the famous $V/R$ cyclical variations 
of Balmer line profiles over years. Examples of Be stars that have about
the velocity separation for the two peaks, according to these authors, 
are $\alpha$\,Col, 25\,Ori, HR\,2787, $\omega$\,Ori, and HR\,2284.

   We began our analysis by determining the wavelengths of the emission
features in the January coaddition spectrum. 
After correcting for the star's radial velocity of +9 km\,s$^{-1}$, 
determined by TTW73, we measured the centroid velocities of the features. 
We found a mean radial velocity of +6${\pm 5}$ km\,s$^{-1}$
for the emisson features we could reliably measure. This value
agrees well with our adopted velocity for the star. 

  Next, we identified lines by computing synthesized LTE 
spectra using the {\it SYNSPEC} and {\it CIRCUS} codes. We utilized
{\it CIRCUS} to simulate the effects of a putative circumstellar ``cloud"
located outside the direct line of sight to the star. The contribution it
makes to emission can be computed for any arbitrary doppler velocity. 
We performed trial computations for models with several temperatures and 
obtained a good match with the line opacity spectrum computed with $T$ 
= 10,000\,K. The identifications, the excitations of the lower
atomic levels, and measured and 
computed equivalent widths of the $V$ and $R$ emission lobes are indicated 
in Table\,1. Almost all the lines in the spectrum are Fe\,II lines.
Typically these have excitations of a few eV, as noted in the table.
Equivalent widths are not given in the table for those cases for which line 
strengths are too weak to make a reliable measurement, or for which two 
or more pairs of $V,$ $R$ lobes are present because of contributions 
from neighboring lines.  The He\,I lines are listed twice, once for the 
January and February observing runs. A comparison of these entries
demonstrates that He\,I line strengths increased during this interval.

  Figure\,\ref{vrhrn} shows a montage of profiles of the He\,I 
5876 \AA\ and 6678 \AA\ lines, as well as a selected group of metallic
Fe\,II lines for both January and February spectra.\footnote{
We can add from an examination of a few archival spectra obtained 
with the ESO 3.6-m. Coude Echelle Spectrograph, that certain weak lines 
(e.g., Fe\,II 4309\,\AA) have emission profiles similar to those in our 
montage.} In aligning the spectra in our figure, we corrected the 
for the star's radial velocity. The vertical lines in the figure denote
the centroid negative and positive velocities of the $V$ and $R$ peaks. 

A comparison of the shapes of the emission lines in Fig.\,\ref{vrhrn} 
reveals that they are virtually identical for lines of any ion. 
However, small differences in 
{\it strengths} are evident from epoch to epoch. First, the February 
profiles show peaks that are closer together than the January profiles 
(${\pm 115}$ km\,s$^{-1}$, instead of ${\pm 102}$ km\,s$^{-1}$). 
Second, the February metallic-line 
emissions represent a $\sim$10\% strengthening over 
the previous month. An additional property of the red He\,I lines in both 
datasets is that the ratio of $\lambda$5876 to $\lambda$6678 emission 
equivalent widths is close to 2. 
Because the ratio of the atomic $gf$ values for these two lines is 2.5, 
this value indicates that the He\,I lines are nearly optically thin
However, the decrease of the ratio in our February data, along with
a small strengthening of the Fe\,II lines, could indicate some 
increase in the column density of the disk during this period. 
According to our simulations below, the metallic lines are
optically thin ($\tau$ $\le$ 0.1).

\subsubsection{Analysis of the emission spectrum}
\label{emissp}

  The double-lobed structure we observe in the optical spectrum of \hd~ 
is the classical shape of an optically thin disk viewed nearly edge-on. 
Therefore, we have assumed that this is the correct viewing geometry
in evaluating the equivalent widths of these lobes. We measured 
equivalent widths by assuming that the lobes have gaussian cores and
symmetric extended wings. We measured their half equivalent widths 
by measuring their contributions between wavelengths defining the central
core of the lobe and the point where the wing merged with the stellar
continuum, typically 50 km\,s$^{-1}$ from the centroid of the core. 
In column 4 of Table\,1 we list the sums of these values for 
the $V$ and $R$ lobes. These are the means of the equivalent widths 
of these two features. The last column of the table is the computed
equivalent width obtained from the T = 9,800\,K model defined below. 
This value is computed by subtracting the narrow contribution of the 
disk component produced by {\it CIRCUS} from the rotationally broadened
photospheric line given by {\it SYNSPEC.} Since the observed profiles 
in Fig.\,\ref{vrhrn} are the same {\it inter alia,} to within our
errors of measurement, we can evaluate attributes such as the 
strengths of their emission lobes straightforwardly.
   
  The determination of specific parameters for our homogenous disk
model proceeded according to the following steps: 

\noindent {\it a) Disk Temperature} 
Lines of neutral helium and Fe$^{1+}$ 
co-exist in the optically thin regime only in a narrow region of
temperature. Therefore, we have exploited this fact to use the ratio 
of the strengths of these lines to determine T$_{disk}$.
Using {\it CIRCUS} models, we found that the number of helium atoms 
populating their 21\,eV levels drops quickly from a moderate finite 
value at 11,000\,K to effectively zero at 9,300\,K (N$_{e}$ = 10$^{11}$ 
cm$^{-3}$), assuming normal chemical abundances.
The single temperature that best fits the observed ratios in Table\,1 is
9,800${\pm 200}$\,K. The error bars on this figure are determined as much 
by our lack of precise knowledge of the volumetric electron density as
by photometric or measurement errors.\footnote{We
believe that the poor fit for the Fe\,II 6318 \AA\ may be due to
inaccurate atomic data for this line.} 

\noindent {\it b) Column density:} 
The measured equivalent width ratio of the
helium lines EW($\lambda$5876)/EW($\lambda$6678) $\approx$ 2.0 provides 
a rough estimate of the optical depth in these lines. Having estimated
the disk temperature from item {\it a}, we computed models with varying 
mean column densities until we found a value that produced a match 
with this equivalent width ratio. We found the most probable value is
3$_{-1}$$\hspace*{-.16in}^{+2}$$\hspace*{.0in}\times$10$^{22}$\,cm$^{-2}$.
The errors here are based on the temperature errors given above, combined
with an assumed ${\pm 5}$\% error in the measured equivalent width ratio. 
Note in this region of temperature-depth space, the propagation of the
errors is asymmetric around the mean value. For this column density,
temperature, and microturbulence (see {\it c}) the Fe\,II $\lambda$5018
and $\lambda$4923 lines are marginally optically thick, having 
$\tau_{\rm line}$ = 2 and 1, respectively.  The optical thickness of
the strongest helium line, $\lambda$5876, is 0.2.

\noindent {\it c) Microturbulence:} This value was determined by 
matching the ratios of Fe\,II lines having the same excitations but 
comparatively low and high $gf$ values, e.g. $\lambda$4923 and 
$\lambda$5018. The value determined from these ratios in our line 
syntheses is $\xi$ = 10 ${\pm 3}$ km\,s$^{-1}$.

\noindent {\it d) Projected area:} 
Finally, the strength of absorption or emission lines sets the projected
star disk in natural units of a star area, $\pi R_{*}^{2}$.  Note that
all other quantities outlined in {\it a-c} utilize line strength {\it 
ratios.} The projected area we determine from our {\it CIRCUS} models 
is the value used to match the {\it absolute} computed strengths to 
the observed ones (columns 4 and 5 in Table\,1). The computed area is
100\,${\pm 10}$R$_{*}^{2}$ (the error bars are internal). Since the disk
is almost in the thin regime, and we may reasonably expect it to be
at least roughly axisymmetric, considerable degeneracy exists for estimates
of disk area and the column density. While the density in particular is 
not well determined, the volume estimate is more secure. 
Assuming a stellar radius of 7R$_{\odot}$, the equivalent disk volume
and mass for the homogeneous model parameters are $\sim$10$^{48}$ 
cm$^{3}$ and $\sim$10$^{-9}$M$_{\odot}$. This is a typical mass for
a well-developed disk of a classical Be star disk. 

  As a clarification, we should note that the solution described above 
represents a volume-weighted mean for the sector of the assumed homogeneous 
disk we have considered.  A more realistic approach to modeling the disk 
segment in the sky plane might take into account that the disk temperature 
and hence the He\,I/Fe\,II emission should vary with distance from the 
star. To evaluate the effects of a temperature gradient, we fit several 
lines in Table\,1 with a {\it two-temperature} component solution, that
is with two distinct but equal volumes having the same column densities 
and arbitrarily chosen temperatures of 10,000\,K and 9,000\,K. 
(Because the radius-temperature relation is unknown, distances of these
two disk components along the line of sight are unknown.) We also scaled 
the areas of the two equal-area components until they fit the  He\,I and 
Fe\,II lines in the table to the same precision as our first model.
We found a good match when the two subareas were 37 stellar areas each. 
This is not to say that the two-parameter model is a better fit to
the equivalent width data than our single parameter model given above.
Rather, it demonstrates only that the added sophistication reduces the
area by only 26\%. We can conclude that the
large areas in our solutions do not seem to be strongly determined 
by radial temperature gradients through the disk.
  
  A second important point to come out of our analysis relies on the
kinematical argument already alluded to in our reference to the work
by Hanuschik (1988). In the case of \hd110, 
the velocity separation of the $V$ and $R$ lobes 
is Keplerian, the region where they are formed is roughly 1.06\,A.U., 
assuming a stellar mass of 12 M$_{\odot}$. 
This leads us to evaluate whether the temperature
we have derived, 9,800\,K, is reasonable for the disk of a B1e star.
By way of comparison, we take consider the theoretical analysis of this
star's disk by Millar \& Marlborough (1998). In their analysis of the 
radiative energy losses and gains, these authors found a density-weighted 
mean temperature of 10,800\,K.  This determination is in 
reasonable agreement with the value of 9,500$\pm{1000}$\,K that Hony et 
al. (2000) derived in their analysis of the bound-free absorption edge of 
the hydrogenic Humphreys jump in infrared spectra. However, according to 
these authors this emission is formed 
primarily within a few stellar radii from the surface of \gam.~ 
Millar \& Marlborough's results indicate that the disk temperature
decreases slowly with stellar distance. Thus, an extrapolation of their
model out to $\approx$ 1\,A.U. ($\sim$30R$_{*}$) would certainly predict 
a lower value than we have found, even though \gam~ is likely to be the 
hotter of the two stars. In this event, it may be that nonradiative 
heating is required to maintain the temperature of the \hd~ disk.

  Third, we note that our disk dimensions, especially an extension out
of the plane, do not necessarily agree with those of other well studied 
stellar disk systems. In particular,
Be disks originating from decretion or protostellar collapse have
densities that are strongly confined to the equatorial plane, and their
contours flare outward from the plane. The well studied disk of \gam~ 
is a case in point. As mentioned above, observations show that a typical 
electron density at a typical point of this disk is $\sim$10$^{12}$ cm$^{-3}$, 
and the column density in the poloidal direction is about $10^{23}$ cm$^{-2}$ 
(Millar \& Marlborough 1998). Interferometric observations in
H$\alpha$ light find that hydrogen is ionized out to a radius of about 
6R$_{*}$ (Quirrenbach et al. 1997). A comparison with this detailed 
description does not allow much leeway to extend the disk of \hd~ 
in the radial direction.  Given the optical thinness of all the lines,
including of He\,I, the only effective 
way to realize a large projected area is to invoke a geometry in which 
the disk flares outward from the plane by several R$_{*}$. We trust
that this apparent peculiarity will turn out to be a clue to resolving
how this star's disk is maintained.

  As a fourth consideration, the question arises whether similar
emission features might also be present in the \gam~ spectrum. In fact,
Bohlin (1970), Slettebak (1982), and evidently TTW73 have already noted 
Fe\,II emission lines, including the first two Fe\,II entries in our table,
and unidentified ``chromospheric" lines, some of which overlap those in
Table\,1, were identified long ago by Heger 
(1922).  Bohlin's paper also referenced the identification of iron emission 
lines by Baldwin (1942) and Cowley \& Marlborough (1968) using
moderate-dispersion photographic plates. In the intervening time, 
it appears that neither the strengths nor the shapes of these features 
have been studied in the \gam~ spectrum. However, from our analysis
of the \hd~ optical spectrum, we can predict that 
future observations of the \gam~ spectrum will disclose numerous 
emission lines with plateau-shaped profiles. Because \gam~ presents an 
intermediate observing angle, it is not clear that its emission line 
profiles will include a well-resolved lobe structure, although this may still
turn out to be the case. In any case, the fact that the lines are present in
the \gam~ spectrum confirms our geometrical picture that the lobe-structure 
in the \hd~ spectrum is due to Keplerian rotation of the disk.

\subsection{The absorption lines of \hd }
\label{brabs}

\subsubsection{Description}

  The only optical line entirely in absorption in our spectra is 
He\,I 4471 \AA.~ In Be stars the nearby Mg\,II 4481 \AA\ 
line is generally easily identifiable, despite its position in the red 
wing of the He\,I feature. In our spectrum the Mg\,II feature has been 
overwhelmed by the particularly strong absorption red wing of He\,I. 
The wings of this line are broadened symmetrically to $\approx$${\pm 1000}$ 
km\,s$^{-1}$. This smearing is shown for the He\,I  $\lambda$4471, 
$\lambda$5876, and $\lambda$6678 lines in Fig.\,\ref{heisiv}. This same 
characteristic is shared by first three Balmer members. 

  This broadening is unlikely to be instrumentally induced for at least two
reasons. First, the spectrograph has been used for years to observe spectra
of other rapidly rotating AB stars, and the recovered profiles are normal
and are in agreement with other instruments. For example, Balona \& James
(2002) have studied the line profiles of several of the He\,I and Balmer
lines in question, including $\lambda$4471. The wings and associated nearby
continua exhibit no abnormalities. Likewise the profiles in our January and
February spectra show no systematic differences from one another.
Second, we may assess the reliability of the continuum and echelle blaze
functions by comparing the echelle order ($m$ = 45) containing the
$\lambda$4471 line with the instrumental continua of the two neighboring
echelle orders.  In Fig.\,\ref{heisiv} this spectrum, and implicitly the
associated internal errors, are labeled by the annotation ``Continuum." 
The key point of this comparison is that fluxes of the auxiliary spectra 
show no, or at most a minor deviation, from flatness across the central 
region of the blazes. 

  We have also considered spectra of other objects taken during the
January and February runs with the same instrumental configuration.  
In Fig.\,\ref{heisiv} we also exhibit the spectrum of the order containing
$\lambda$4471 for the roAp star HR1234 observed during the January run.
We note that an A3 star was observed during the February run. Once again,
the orders containing this line and the red helium lines were flat whereas
the spectra of HD\,110432, reduced in exactly the same way, showed the 
characteristic ${\pm 1000}$ km\,s$^{-1}$ broadening. 

  It is important to note that the core depth $\lambda$4471 in our 
spectra is 10\,\%. According to profiles of the same line in the spectra 
of $\gamma$\,Cas (Chauville et al. 2001) and other rapidly rotating B0-2V
stars found in the UVES Paranal spectral atlas (Bagnulo et al. 2003),
this is the  nominal depth of this line in spectra of rapidly rotating 
early-type Be stars. Rotational broadening dominates the shape of the 
(weak) far wings in the latter cases, a fact we have confirmed in our
own spectral syntheses of hydrogen and helium lines.
For this reason the absorption lines should be characterized as {\it 
strengthened} as well as anomalously broadened.

 The only other published report of a optical-wavelength helium line
profile is that of Ballereau, Chauville, \& Zorec (1995) atlas of 
$\lambda$4471, $\lambda$4481, and H$\gamma$ profiles of Be stars. 
The $\lambda$4471 of \hd~ in this atlas is somewhat peculiar, 
exhibiting both a sharp line core and triangular wings, and in this
sense it appears unlike other $\lambda$4471 profiles in the atlas.
Nonetheless, it is quite unlike the profile in 2005. It is evident that
the line's shape, if not strength, has changed on a timescale of several 
years.

\subsubsection{The unknown broadening agent }

  In any context the presence of the strong absorption lines broadened 
by more than twice the star's likely rotational velocity presents an
interpretational challenge. These absorption lines cannot arise from the
blended superposition of two components of a double-lined binary because
this circumstance would not produce the substantially strengthened lines 
we observe.  Moreover, spectra of a presumed near edge-on Be-binary 
system should exhibit radial velocity variations within a few
days, and these have not been observed. In this model, lines might be
rendered broad and unresolved at certain phases, but they would not be 
strengthened.

 Because the star's parallax is in agreement with its spectroscopic 
luminosity class, it is hardly possible for a compact star such as a 
white dwarf to dominate the flux contribution of the B star and produce 
Stark broadened wings. However, the profiles in Figs.\,\ref{heisiv} and
\ref{vrhrn}, including H$\gamma$ and the companion profiles of the
February dataset (not shown) indicate that the
line wings end abruptly at a flat continuum some 1100--1300 km\,s$^{-1}$
from line center. In this sense they do not resemble the gradual 
merging into the continuum of pressure-broadened line profiles.
  
  At the referee's suggestion (and for the moment suspending our knowledge
that HD\,110432 occupies a position in the H-R Diagram appropriate to
a $\sim$B1\,IV star and also that the profiles seem to be variable over
an underdetermined long timescale), we have also investigated  
the line profiles of the He\,I lines in the early DB white dwarfs. 
In surveying the literature, we find it is comparatively easy to find
examples of 4471\,\AA\ profiles that are Stark broadened to 
considerable degrees (Wesemael et al. 1993, Liebert et al. 2003).
However, many of these same spectra exhibit no visible He\,I red lines. 
The red lines of two sdB stars observed by Heber \& Edelmann 2004)
exhibit very weak wings.
Among three early DB stars observed by Wolff et al. (2002), only one 
of them has spectral line wings that extend to ${\pm 350}$ km\,s$^{-1}$. 
We can conclude that since the Stark components are 
weak in the red lines of these sdB and DB stars, the excess broadening 
found in the helium and hydrogen lines of our spectra obtained in
2005 is not caused by pressure broadening. This conclusion is likely
to extend to the other broadened lines in the blue spectral region.

  Electron scattering can also broaden lines over a velocity range
characteristic of the temperature in a hot gas. However, this 
mechanism is conservative in the sense of redistributing monochromatic 
flux rather than a strengthening of a line. 
Moreover, for electron scattering to be effective the required 
very high column densities (10$^{24-25}$ cm$^{-2}$) are not present 
near \hd, according to both analyses of our optical emission line 
spectrum and of the gradient of the X-ray continuum (TO01). 

  By process of elimination, we are left with invoking doppler
velocities to explain the excess absorption line widths in optical lines.
However, we have no information at the moment to link such velocities
to other phenomena, including the X-ray emissions. We will continue
to monitor this star's optical spectrum both to confirm this behavior
and to investigate whether the broadening evolves in the coming years.

\section{Conclusions}
\label{concl}

  \hd~ is an unusual Be star, whether studied in the X-ray, ultraviolet,
or visual regimes. Its hard X-ray spectrum is probably thermal (TO01).
As shown in Figs.\,\ref{xrhd} and \ref{xrgam}, its optical light curve for
the year 2002 shows modulations which arguably are similar to those observed 
in \gam.~ This makes \hd~ a fascinating target because it makes its 
association with the highly enigmatic \gam~ possible. Although the far-UV
and UV lines are representative of a rapidly rotating B1 star, like a number
of Be stars its resonance lines show variations on a timescale of a few
days, or perhaps much less. The hydrogen and helium line absorptions of the
{\it Giraffe} spectra obtained in January and February 2005 appear to be 
``almost photospheric." These lines suggests a $\approx$B1 spectral type 
that is consistent with the spectral energy distribution determined by 
Codina et al. (1984). However, the wings of these lines are strengthened 
by a broadening corresponding to $\approx$$\pm{1000}$ km\,s$^{-1}$. The 
cause of this broadening is unknown.

  A number of metallic emission lines chiefly due to Fe\,II 
are present in the green-red spectrum. These features exhibit a 
central plateau flanked by two $V$ and $R$ emission peaks. This same
profile is shared as an inner core feature of the first three members 
of the hydrogen Balmer series and the He\,I $\lambda$5876 and $\lambda$6678
lines.  The $V$ and $R$ peaks of all these emission lines are 
separated by  about 200 km\,s$^{-1}$. Both the separation velocity
and the strengths of the featurers can vary on a timescale of a month. 
We can conclude from {\it i)} the shape of the emission
line profiles, {\it ii)} the equivalent width ratio of the features in 
the two ``red" He\,I lines, and {\it iii)} analogous emissions in the 
\gam~ spectrum that the emission lobes are formed in an optically thin, 
Keplerian disk extending out to about 1\,A.U. from the star. Using this 
basic geometry, we can solve for the parameters of a homogeneous disk. 
Along the sight line of maximum elongation of the projected disk,
the mean disk temperature is close to 9,800\,K, the column density
through this region of the disk is 
roughly 3$\times$10$^{22}$ cm$^{-2}$, and the projected area is some 100 
stellar areas. (A likely small decrease in the He\,I line ratio could indicate
a mean column density of perhaps 4$\times$10$^{22}$ cm$^{-2}$ at this epoch.)
The latter value is quite large and suggests that the 
disk flares from the central disk plane. The computed disk area can be 
reduced by assuming that the disk has an arbitrary temperature gradient.  
At a distance of $\approx$1 A.U. from the star, a gas temperature of 
9800\,K may be difficult to maintain from the star's radiative flux alone. 
Therefore, it is possible that an additional heating source is required.

  The X-ray and optical properties of \hd~ are similar to those of
\gam, a fact that suggests that the production mechanism of the X-rays may
well be the same for these two stars. They are both early-type stars 
near the main sequence with a high rotation rate, even for a Be star.
Their disks are well developed and both exhibit metallic and emission
line spectra. In assessing the X-ray properties, we see that both stars
exhibit signatures of plasma 
with a temperature of about 10$^{8}$\,K and variations over a timescale
of hours that can be characterized, at least with the available light
curve, as chaotic meandering.  We might hazard a
prediction that, as with  \gam,  future high quality X-ray light curves of 
\hd~ will show nearly continuous flaring  and (from interaction with an
extensive disk), and a visible fluorescent Fe K line at 6.4\,keV. Turning 
to our own optical work, first in a ground-based photometric campaign 
in 2002, the star's light curve during this season undergoes modulations 
with an amplitude of at least 3\% that are consistent with a cyclical 
variation of about 130 days. This is consistent with the optical cycles of
\gam, which have cycle lengths of about the same amplitude of 2--3 months.  
In both stars the cycle amplitudes increase from
the $B$ to $V$ filter wavelengths. From an analysis of a spectroscopic 
time series of the red He\,I lines over several nights in early 2005, 
we have also found traveling absorption features migrating
through the profiles at a rate equal to the {\it moving subfeatures} often
visible in the profiles of \gam. Such features are highly unusual in
Be stars, and indeed so far are clearly present only in \gam~ itself.
Their existence is best, though not uniquely, explained by absorptions of
small clouds that are locked into co-rotation by surface magnetic fields. 
This explanation has been bolstered by the recent report of the discovery
of a long-lived periodicity of 1.21 day on \gam, probably caused by
rotational modulation of a magnetic surface structure (SHV06).

   From these observations we suggest that \hd~ is the first new member of
what may be called a ``\gam~ class" of X-ray Be stars, so far consisting
of two stars.  In the case of \gam, the prevailing evidence is that these 
X-rays are produced in the immediate vicinity of the Be star itself, perhaps
by magnetic disk-star interactions (RSH02). Whatever the physical site,
we may begin to evaluate candidate X-ray production mechanisms by
examining the properties of both the X-ray and the star among members 
of this new class. Moreover, the greater the number of stars amassed
in this class, the greater will be the probability of monitoring disk 
changes as they occur, and thus of determining the relevance of the disk 
to the high X-ray emission. We note a paper by Motch et al. (2006) that 
came to our attention as this paper was being completed has identified five
possible $\gamma$\,Cas analogs, including HD\,110432 as the brightest, 
based on their X-ray and optical properties. 

  In assessing differences in phenomenology 
that might be observed in \gam~ and \hd, the aspect angle to the observer 
could play an important role.  According to interferometric studies of \gam~
(e.g., Quirrenbach et al. 1997), the star-disk system is observed from an
intermediate obliquity. Smith et al. (2004) found that its {\it Chandra} 
continuum spectrum is attenuated at longer wavelengths in such a way as
to require a two column-density absorption model to fit it. If the X-rays 
arise from a star-disk interaction, this model fits in well with the 
expectation that these emission would be emitted from two types of sites, 
one primarily in front of the disk along the line of sight and the other 
behind it. If one were to observe $\gamma$\,Cas or an equivalent Be
star from an edge-on orientation, and further, were to assume that most
of the emission occurs outside the disk plane, then one might expect the 
continuum to be fit to a single column-density model. The column length
derived would be high if the active centers were close to the disk plane 
but nearly zero if they were distributed far from the plane. The line
spectrum of H-like light ions and Fe L-shell ions are primarily located
between $\approx$8--17\AA\, more or less intermediate between the wavelength
ranges used in the continuum to derive column absorption properties.
A change in observer aspect should leave the line spectrum in this 
intermediate wavelength range less changed. Interestingly, T001 found 
an intermediate column density of 10$^{22}$ cm$^{-2}.$ 
A high-resolution spectrum should be able to address this question 
further and lead to constraints on the geometry of the formation region. 
Proceeding in the opposite direction, an understanding of disk dynamos 
could lead to an understanding of conditions in the inner disk.

   This paper could not have been written without help from a number 
of colleagues who provided us with information and encouragement to
unravel peculiar properties of this star. Thanks
are due first to Dr. J. Torrej\'on and the editor of the {\it Astronomy \& 
Astrophysics} journal for giving their  permission to utilize 
the light curve figure from the TO01 paper. We also acknowledge a
number of helpful discussions about the properties of this star by
Drs. Alex Fullerton, Steve Howell, Ted Snow, and Janez Zorec. We also
thank Drs. Andrea Torres and Adela Ringuelet for providing us with
unpublished diagrams of the dependence of the He\,II $\lambda$1640
line on rotation.  We gratefully
acknowledge the fruitful efforts of Mr. Francois van Wyk, who carried
out the photometric monitoring on \hd~ in 2002 and of Dr. David Laney
who kindly obtained many spectra of this star during 2005 February.
The quality of this paper was significantly
enhanced from comments by an anonymous referee.

\clearpage

\centerline{\bf Figure Captions }

 \begin{figure}
\vspace*{-0.1in}
 \caption{
  A FUSE spectrum of HD\,110432 and a spectrum of \gam~ shown (dashed
line) for
reference for the wavelength region $\lambda\lambda$1115--40. The light
solid line is a scaled synthetic spectrum of the photosphere modified
by a minor contribution from a warm circumstellar disk. The synthetic spectrum 
has been broadened by a rotational quasi-convolution of 300 km\,s$^{-1}$. 
}
\label{fit11}
 \end{figure}
 \begin{figure}
\vspace*{-0.1in}

 \caption{
  A comparison of spectra in the wavelength region surrounding the 
He\,II $\lambda$1640 line of HD\,110432 (all available 13 IUE/SWP spectra), 
$\gamma$\,Cas (all 31 small-aperture IUE/SWP spectra obtained in 1982), 
and the SYNSPEC-synthesized spectrum.}
\label{he2fit}
 \end{figure}

\begin{figure}
\vspace*{-0.1in}
 \caption{ Panel {\it a} shows the
Cousins V-band light curve of \hd~ during 2002 February--June.
If this variation is a cycle, it has a length of perhaps 130 days.
Panel {\it b} shows a scatter diagram of the V- and B-band variations
for the monitoring shown in panel {\it a.} The reference for
the points in the upper left of panel {\it b} indicates that the
optical variations are slightly larger for the V-band than the
B-band. }
\label{ltcol}
 \end{figure}

 \begin{figure}
\vspace*{-0.1in}
 \caption{
A {\it RXTE} light curve of \gam~ from a 6-hour section
of the {\it RXTE} light curve obtained on 1998 November 25--26; the solid
line is the sine curve from Fig.\,\ref{xrhd}. Its amplitude and period are 
the same as in the first case. Only the mean flux level of this curve
has been modified to fit the \gam~ flux. }
\label{xrgam}
 \end{figure}

 \begin{figure}
\vspace*{-0.1in}
 \caption{
 A 5-hour {\it BeppoSax}
light curve of \hd~ constructed by Torrejon \& Orr (2001). The solid
curve is a 4-hour sine wave fit to the dip in the middle of the
observations. {\it By permission of Astronomy \& Astrophysics.} }
\label{xrhd}
 \end{figure}

\begin{figure}
\vspace*{-0.1in}
\caption{
Sample observations of the C\,IV doublet complex (bottom) and Si\,IV 
$\lambda$1403 (top) for epochs in 1981, 1986, and 1991, respectively; 
the abcissa is 
velocity. The rest positions of the C\,IV doublet are indicated by the
$\lambda_o$ symbols. The IUE/SWP observing sequence numbers are indicated. 
Dashed and dotted arrows indicate the positions of the central minima 
discussed in the text.}
\label{civ}
 \end{figure}

\begin{figure}
\vspace*{-0.1in}
 \caption{
A comparison of three highly broadened neutral helium lines from our
January 2005  SAAO {\it Giraffe} observations, offset for clarity. Vertical 
dotted lines indicate the approximate velocity limits of the wings of the 
lines. The nearly ``Continuum" lines at the bottom are the superimposed
spectra of the two neighboring echelle orders from the same observations.
Note that the red wing of the $\lambda$5876 spectrum has been truncated
by the edge of the blaze. The spike at the red edge of the $\lambda$6678 
spectrum (scaled by a factor of two) is a grating-induced Wood's anomaly. 
The plot at the bottom is the spectrum of the wavelength region near
$\lambda$4471 of the roAp star HR 1234. This spectrum was obtained
during the January run when \hd~ was observed.
}
\label{heisiv}
 \end{figure}

 \begin{figure}
\vspace*{-0.1in}
 \caption{
Grayscale difference spectra of the He\,I 6678 \AA\ line of
\hd~ on the dates indicated. Starting times are 23:00 U.T. and 23:15 U.T.,
respectively. Limits of ${\pm 300}$ km\,s$^{-1}$ on the line profile are
indicated as the approximate extent of the rotationally
broadened component of the absorption profile. }
\label{hdmsf}
 \end{figure}

\begin{figure}
\vspace*{-0.1in}
 \caption{
Montage of He\,I and Fe\,II line emission profiles from the mean spectra
of the January and February, 2005 monitoring campaigns.  The dashed and
dotted vertical lines denote the mean separation of the $V$ and $R$
emission lobes for the January and February datasets. }
\label{vrhrn}
 \end{figure}

\clearpage


\begin{table}[ht!]
\begin{center}
\caption{\label{}\centerline{Emission Line Identifications and Strengths }}
\centerline{~}
\begin{tabular}{lrrcc}  \hline\hline

Ion &  Wavelength (vac.) & $\chi$ (eV) &  Observed EW (m\AA) & Computed EW (m\AA) \\
\hline \hline

~S II  &  4524.68-.94 & 15.5 &                    &        \\
Fe II  &  4555.89     &  2.8 &                    &         \\
Fe II  &  4583.84     &  2.8 & -0.182$\pm{.032}$  & -0.182  \\
Fe II  &  4629.34     &  2.8 & -0.076$\pm{.011}$  & -0.058 \\
Fe II  &  4666.76     &  2.8 & -0.080$\pm{.015}$  & -0.056 \\
Fe II  &  4923.93     &  2.9 & -0.101$\pm{.004}$  & -0.103 \\
Fe II  &  5018.44     &  2.9 & -0.252$\pm{.026}$  & -0.312 \\
Fe II  &  5169.03     &  2.9 & -0.264$\pm{.008}$  & -0.258 \\
Fe II  &  5197.58     &  3.3 & -0.150$\pm{.007}$  & -0.137 \\
Fe II  &  5234.62     &  3.2 &                    &        \\
Fe II  &  5276.60     &  3.2 & -0.164$\pm{.028}$  & -0.190 \\
Fe II  &  5284.07     & 10.5 &                    &        \\
Fe II  &  5316.62     &  3.2 & -0.236$\pm{.030}$  & -0.217 \\
~S II   &  5362.87     &  3.2 & -0.122$\pm{.002}$  & -0.090 \\
Fe II  &  5534.89     & 10.5 &                    &        \\
~He I   &  5875.62     & 21.1 & -0.304$\pm{.004}$  & -0.300 \\
~He I (Feb.)  &   5876.62     & 21.1 & -0.361$\pm{.004}$  &        \\
Fe II  &  6247.55     &  3.9 &                    &        \\
Fe II  &  6317.98     &  5.5 & -0.161$\pm{.051}$  & -0.043 \\
Si II  &  6347.11     &  8.1 &                    &        \\
Fe II  &  6456.38     &  3.9 & -0.145$\pm{.024}$  & -0.118 \\
He I   &  6678.15     & 21.2 & -0.146$\pm{.003}$  & -0.150 \\
He I (Feb.)   &  6678.15     & 21.2 & -0.199$\pm{.013}$  &        \\
\hline
\end{tabular}
\end{center}
\end{table}

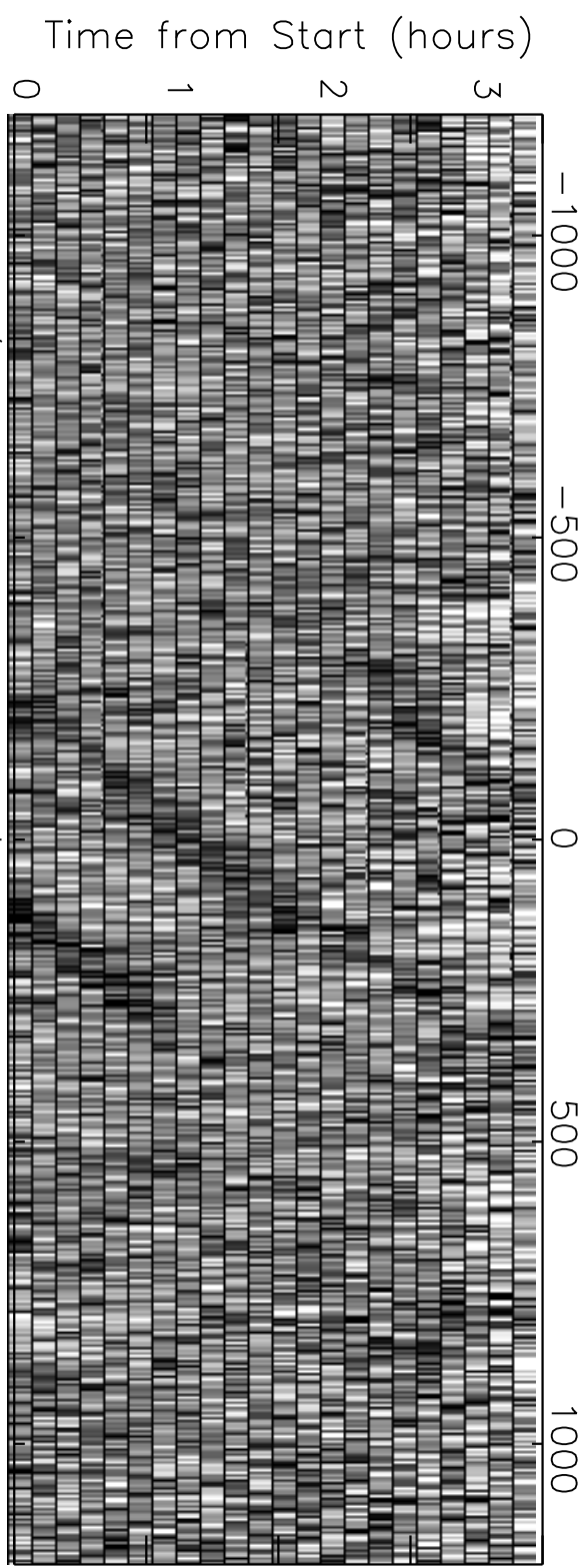
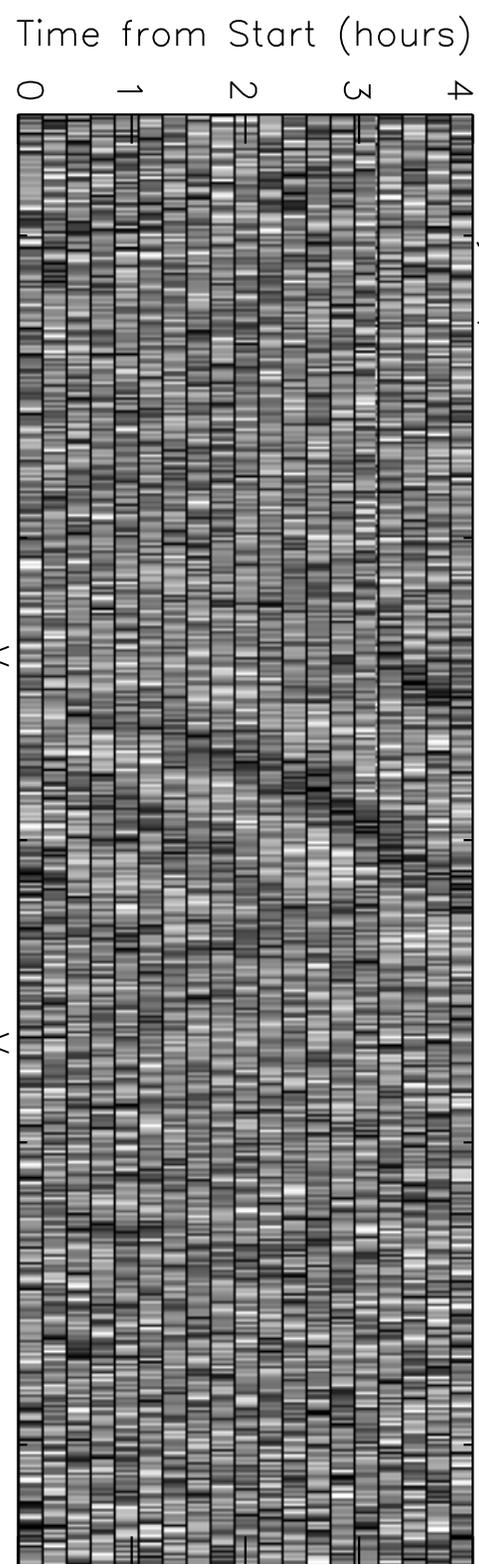
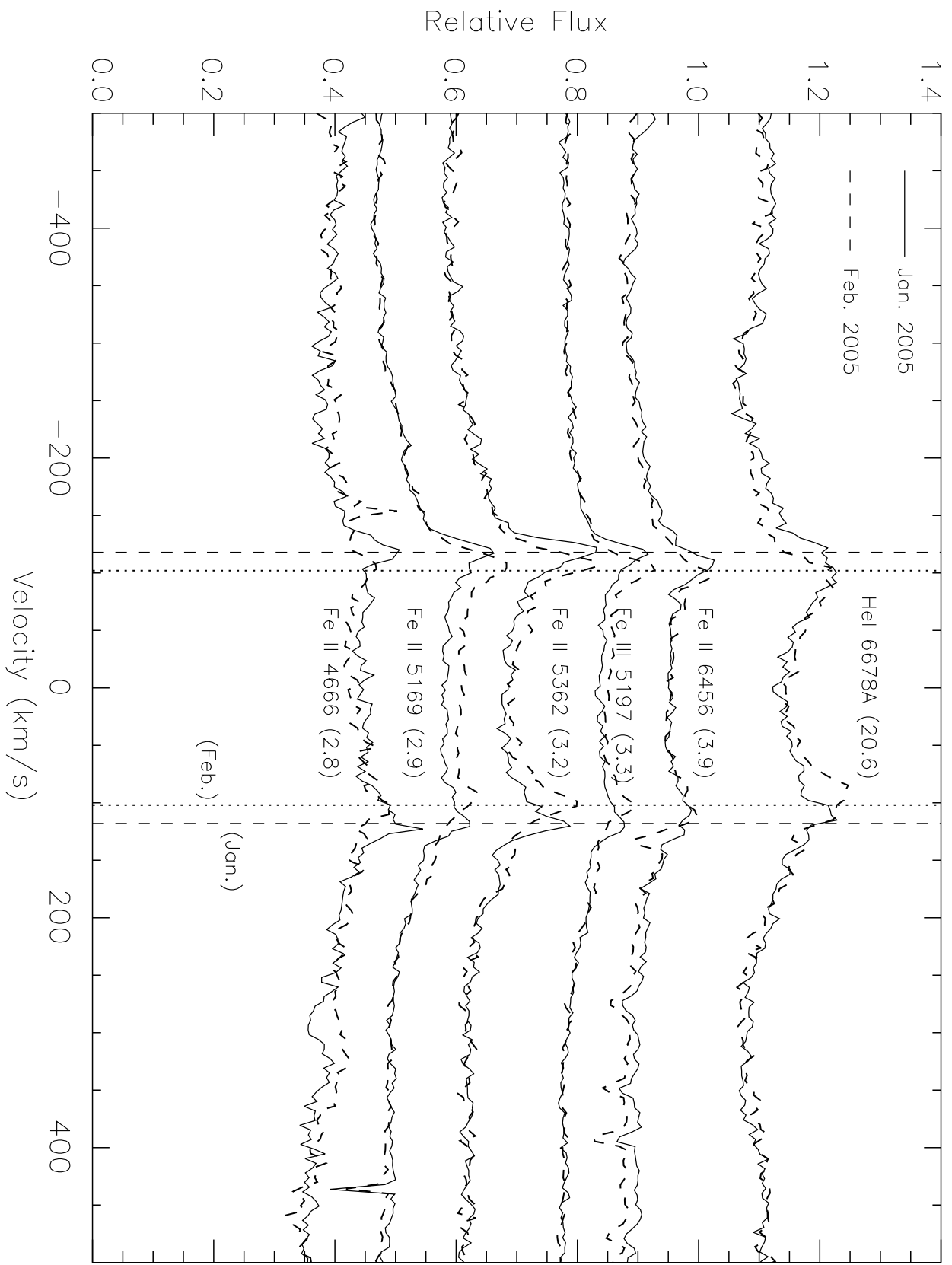
\end{document}